# Designing band gap of graphene by B and N dopant atoms

Pooja Rani and V.K. Jindal[1]

Department of Physics, Panjab University, Chandigarh-160014, India

*Ab-initio* calculations have been performed to study the geometry and electronic structure of boron (B) and nitrogen (N) doped graphene sheet. The effect of doping has been investigated by varying the concentrations of dopants from 2 % (one atom of the dopant in 50 host atoms) to 12 % (six dopant atoms in 50 atoms host atoms) and also by considering different doping sites for the same concentration of substitutional doping. All the calculations have been performed by using VASP (Vienna Ab-initio Simulation Package) based on density functional theory. By B and N doping p-type and n-type doping is induced respectively in the graphene sheet. While the planar structure of the graphene sheet remains unaffected on doping, the electronic properties change from semimetal to semiconductor with increasing number of dopants. It has been observed that isomers formed differ significantly in the stability, bond length and band gap introduced. The band gap is maximum when dopants are placed at same sublattice points of graphene due to combined effect of symmetry breaking of sub lattices and the band gap is closed when dopants are placed at adjacent positions (alternate sublattice positions). These interesting results provide the possibility of tuning the band gap of graphene as required and its application in electronic devices such as replacements to Pt based catalysts in Polymer Electrolytic Fuel Cell (PEFC).

## INTRODUCTION

Graphene is the name given to a single layer of graphite, made up of $sp^2$ hybridized carbon atoms arranged in a honeycomb lattice, consisting of two interpenetrating triangular sub-lattices A and B (Fig. 1) and is a basic building block for carbon allotropes of other dimensionalities like fullerenes and carbon nanotubes. Though it was realized in 1991 that carbon nanotubes were formed by rolling a 2D graphene sheet or a single layer from 3D graphitic crystal, the isolation of graphene sheet was not done till 2004. Since its successful experimental fabrication in 2004 [1], it has attracted enormous interest both from experimentalists and theoreticians. Its unique properties like half integer quantum Hall effect, high charge carrier mobility due to linear dispersion at the so called Dirac point and ballistic transport over long distances, finite conductivity at zero carrier concentration [2], make it an excellent candidate for the next generation of electronics by overcoming silicon-based electronics limitations [3]. A pristine graphene layer is however a zero gap semiconductor (or semimetal) with a point like Fermi surface. Some reviews on the properties of graphene have appeared in the literature, e.g. by Castro Neto et al [4] and Cooper et al [5] which describe pure graphene in detail.

---

[1] Author with whom correspondence be made, Email: jindal@pu.ac.in, present address: Department of Theoretical Chemistry, Technical University, D-10623 Berlin, Germany.

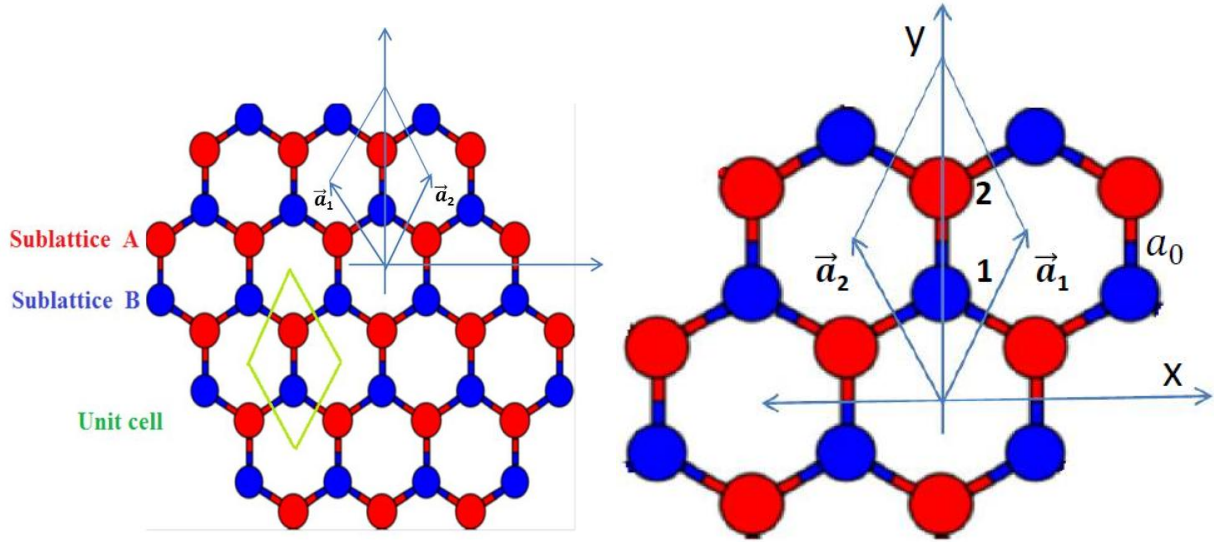

**Fig1. A schematic presentation of graphene sheet. Each Bravais lattice unit cell includes two nonequivalent sites, which are denoted by A and B. A blow up of Unit Cell is shown separately, $a_1$ and $a_2$ are the primitive vectors.**

Pure graphene, though extremely interesting, suffers from zero band gap fixation which makes it uninteresting from device application point of view. The development of graphene based electronics depends on our ability to open a tunable band gap. Various approaches have been developed to fabricate high-performance graphene devices by engineering their band gaps so as to improve their semiconducting properties. One of the approaches is evidently to choose N and B and substitute them to replace C atoms to form 'carbon alloys'. The atomic masses of these dopants is closest to Carbon which would seemingly be acceptable for carbon lattices to adjust to, and at the same time altering significantly the electronic properties of the host material because of electron rich and electron deficient nature of N and B atoms. Experimental and theoretical studies on graphene doping, do show this possibility of making p-type and n-type semiconducting graphene. In fact studies by different means like doping with heteroatoms [6, 7], chemical functionalization [8] applying electric fields and depositing graphene on substrates like SiC, SiO2) [9] have shown this possibility. Band gap opening using p-type doping using Al , Boron, $NO_2$, $NH_3$ $H_2O$, F4-TCNQ and n-type doping using N, alkali metals has been studied in the past [7,10-14]. The observation has been that the dopant atoms can modify the electronic band structure of graphene, and open up an energy gap between the valence and conduction band. Recently, an *ab-initio* study of the different dopant interactions in the host graphene has also been reported which provides useful information on the behavior of dopants [15]. While Lherbie et. al [10] studied the charge mobilities and conductivity of the system by doping graphene with different concentrations of B and N impurities ( upto 4%) randomly ( without studying the band gap), Wu et. al [11] studied the band gap opening in graphene by only single atom doping of B and N. Despite some work by Y. Fan et. al [16] to tune band gap by choosing bi-layers of differently doped graphene, and X. Fan et. al [17] have primarily focussed on Boron-Nitride combination as a dopant and is changing the concentration by taking a larger number of host atoms. Manna et. al [18] have shown that the patching of graphene and h-BN sheet with semiconducting and/or insulating BxNy (Cz) nanodomains of various geometrical shapes and sizes (h-BN sheet) can significantly change the electronic and magnetic properties. Despite all such work, a systematic study of exact role of concentration and position of dopant atoms in modulating the band gap has still not appeared in the literature.

In the present paper we have made an effort to present a systematic study of effect of substitutional doping of boron and nitrogen in the graphene sheet by slowly increasing the concentration of doping and also considering the different isomers of same doped configuration. We choose B and N atoms both for our study, individually at this moment as dopants due to their nearly similar size to that of carbon and because of their electron deficient and electron rich character respectively, and deferring for the present using combination of B and N in host of C atoms. In the following, we outline computational details for our approach of using density functional theory and describe the results for B and N doping in section 3. These results are discussed and finally concluded in Section 4.

## 2. COMPUTATIONAL DETAILS

Graphene is known to relax in 2-D honeycomb structure (Fig. 1) and the B and N doped graphene will be assumed to have similar structure, unless violated by energy minimization considerations. To do this analysis, geometry optimizations and electronic structure calculations have been performed by using the VASP (Vienna Ab-initio Simulation Package) [19, 20] code based on density functional theory (DFT). The approach is based on an iterative solution of the Kohn-Sham equation [21] of the density function theory in a plane-wave set with the projector-augmented wave pseudopotentials. In our calculations, the Perdew-Burke-Ernzerhof (PBE) [22] exchange-correlation (XC) functional of the generalized gradient approximation (GGA) is adopted. The plane-wave cutoff energy was set to 400 eV. The optimizations of the lattice constants and the atomic coordinates are made by the minimization of the total energy. The $5 \times 5$ supercell (consisting of 50 atoms) have been used to simulate the isolated sheet and the sheets are separated by larger than 12 Å along the perpendicular direction to avoid interlayer interactions. The Monkhorst-Pack scheme is used for sampling the Brillouin zone. In the calculations, the structures are fully relaxed with a Gamma–centred $7 \times 7 \times 1$ k-mesh. During all of the calculation processes, except for the band determination, the partial occupancies were treated using the tetrahedron methodology with Blöchl corrections [23]. For band calculation, partial occupancies for each wavefunction were determined using the Gaussian smearing method with a smearing of 0.01 eV. For geometry optimizations, all the internal coordinates were relaxed until the Hellmann-Feynman forces were less than 0.005 Å.

## 3. RESULTS AND DISCUSSION

First of all a pure graphene sheet was fully optimized, including its lattice constant, which was found to be 2.45 Å slightly less than the experimental value of 2.46 Å and the resulting C-C bond length of pure graphene is 1.41 Å which in agreement with previous results [24].The lattice constants, $\mathbf{a}_1$ and $\mathbf{a}_2$, (refer to Fig. 1) are expressed in Cartesian coordinates as

$\mathbf{a}_1 = a_o/2(3, \sqrt{3})$

$\mathbf{a}_2 = a_o/2(3, -\sqrt{3})$

where $a_o$ is interatomic distance or C-C bond length and has been found to be close to 1.42 Å..

Then the band structure of pure graphene was calculated and presented in Fig. 2, which is also found to be in agreement with the literature [2].

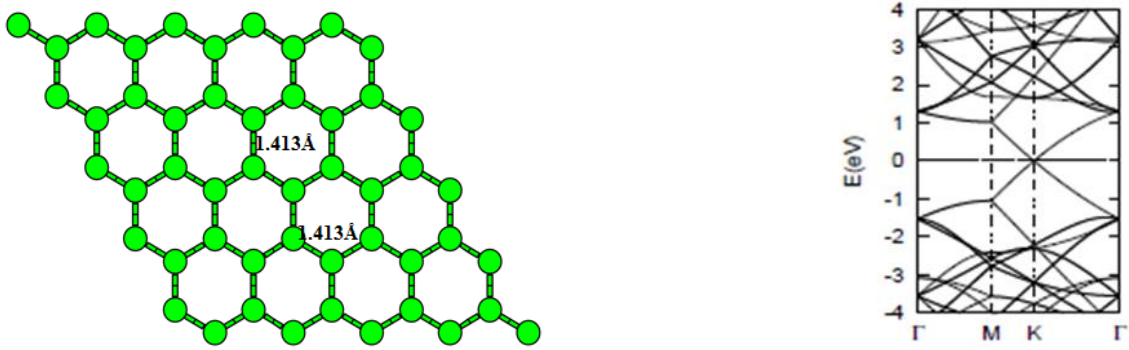

**Fig.2  Optimized geometry and band structure of pure graphene sheet**

Subsequently, pure graphene is doped with increasing concentration of boron and nitrogen atoms. The relaxed lattice constant of doped graphene increases in case of B and decreases in case of N with increase in number of doped atoms because the covalent radius of boron is larger than carbon atom and that of nitrogen is less than carbon and this is consistent with the earlier results as summarized in the Introduction.

We study six B-doped as well as N-doped configurations with 2%, 4%, 6%, 8%, 10%, 12% concentration of doping.

## 3.1 Boron doping

As pointed out earlier, Boron atom is likely to adjust to surrounding C atoms of the host. Therefore when graphene sheet is doped with one boron atom, the boron atom also undergoes $sp^2$ hybridization and due to the nearly same size of C and B, no significant distortion in 2-D structure of graphene is expected, except for change in adjoining bond length. As a result, a bond length is found to expand to 1.48 **Å**. Using the computational procedure as stated above, the electronic properties, especially the band structure is possible to calculate. Of interest are the K-points along special directions of Brillouin zone and we find that, the linear dispersion near the Dirac point is not destroyed. However, due to electron deficient character of boron, Fermi level shifts significantly by about 0.7 eV below the Dirac point resulting into a p-type doping. Another important observation can be made is regarding breaking of the symmetry of two graphene sub-lattices because of introduction of B atom, now introducing an energy gap of 0.14 eV around the Dirac point. This changes the behavior of graphene from semimetal to conductor. The charge transfer was also calculated using Bader charge analysis, which shows 1.79e charge transfer from boron to carbon. The results are in agreement with earlier calculations [11]. Some data concerning a single atom doping by B is presented in table 1.

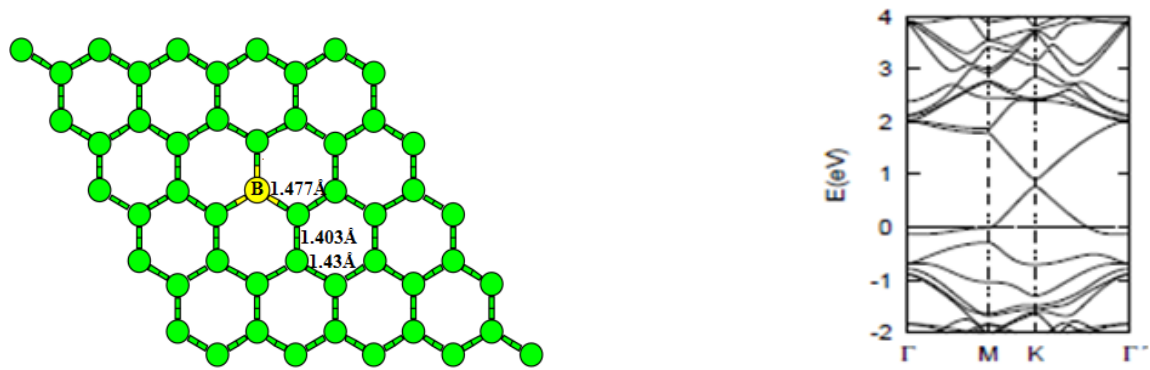

**Fig.3 Optimized geometry and band structures of single boron atom doped graphene sheet**

Table 1. Some parameters affecting single atom doping by Boron atom

| Parameter | $d_{B-C}$ (Å) | Band gap (eV) | Charge Transfer(e) |
|-----------|---------------|---------------|--------------------|
| Our Work  | 1.48          | 0.14          | -1.79              |
| Ref.11    | 1.49          | 0.14          | -184               |

Once we find a satisfactory reproduction of the results as obtained for pure graphene and single atom doped Boronated graphene, we carry out our calculations with increasing number of dopants and also taking into account different sites of doping for the same concentration. It has been found that several isomers of same atomic concentrations doped graphene become possible. These isomers differ significantly in cohesive energy, bond length and band gap. We present the optimized structure as well as the electronic band structure of various isomers of obtained by doping increasing number of Boron doped atoms in Figs. 4-8.

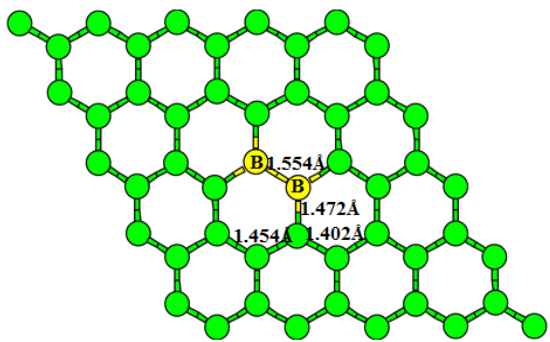
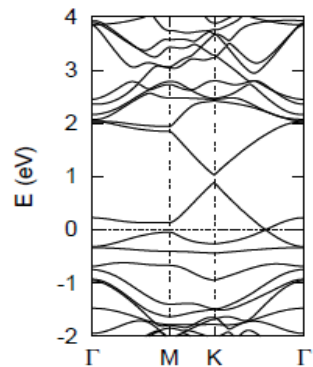

Fig. 4(a)

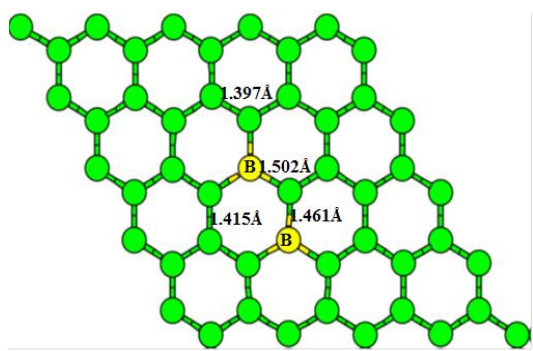
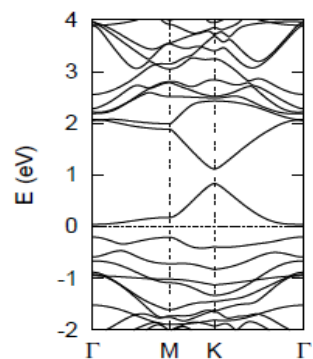

Fig. 4 (b)

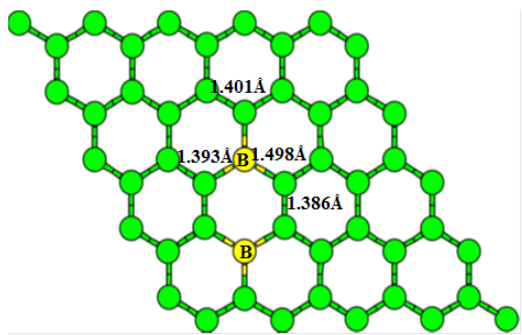
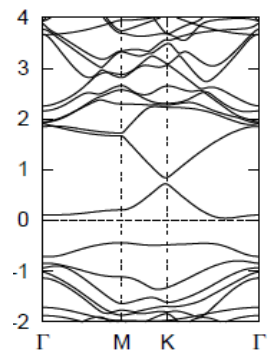

Fig. 4 (c)

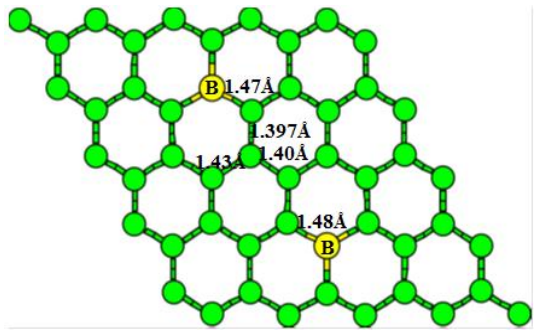 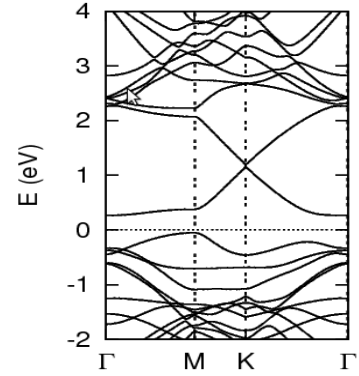

Fig. 4 (d)

**Fig.4  Some optimized geometries and band structures of two B atom doped graphene sheet**

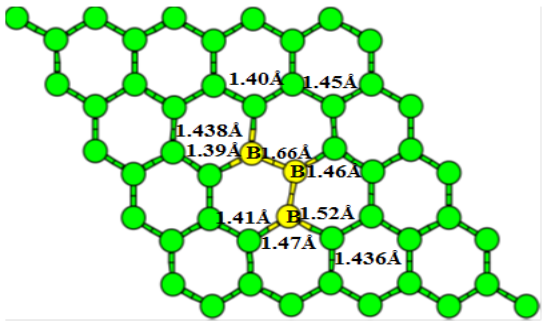 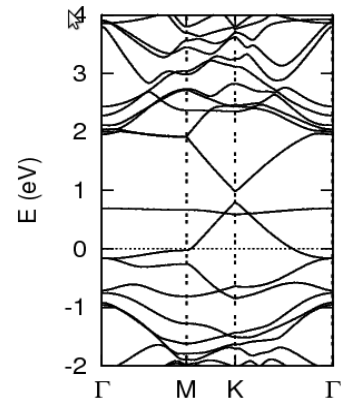

Fig. 5 (a)

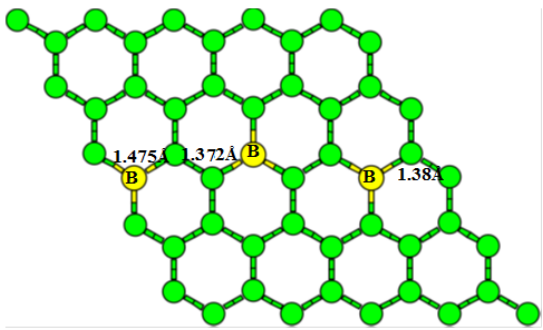 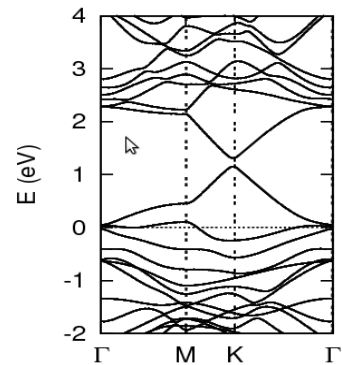

Fig. 5 (b)

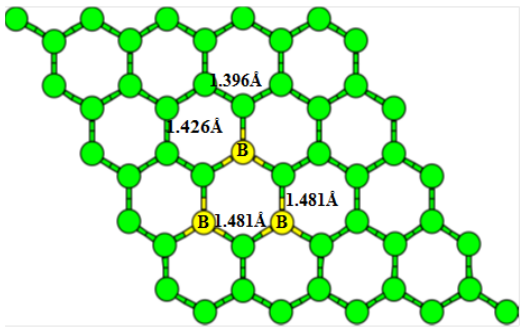 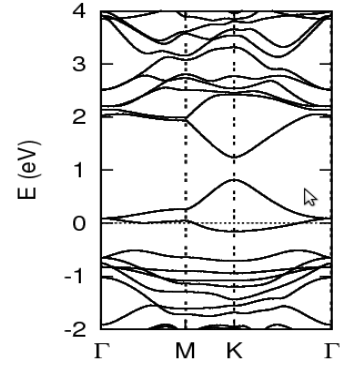

Fig. 5(c)

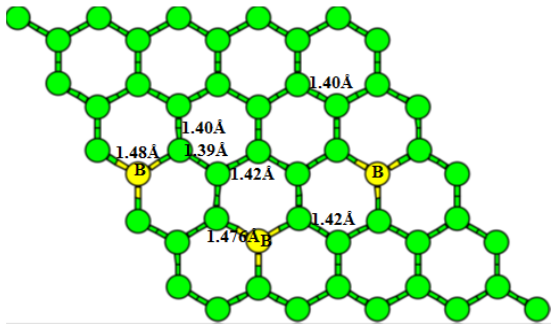 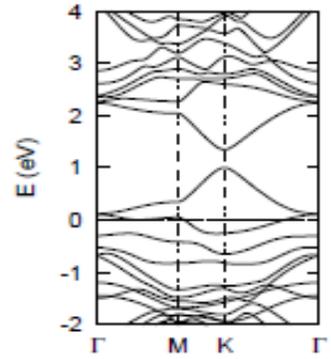

Fig. 5 (d)

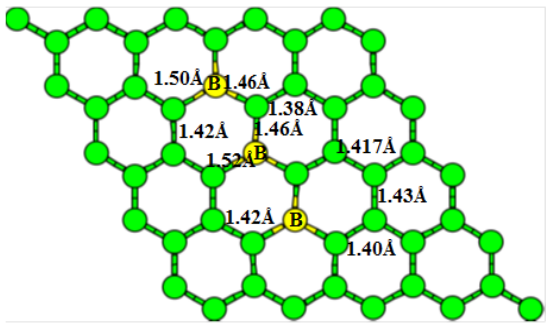 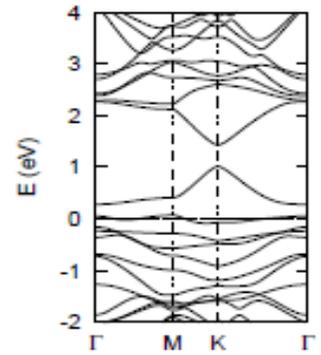

Fig. 5 (e)

**Fig.5  Some optimized geometries and band structures of three B atom doped graphene sheet**

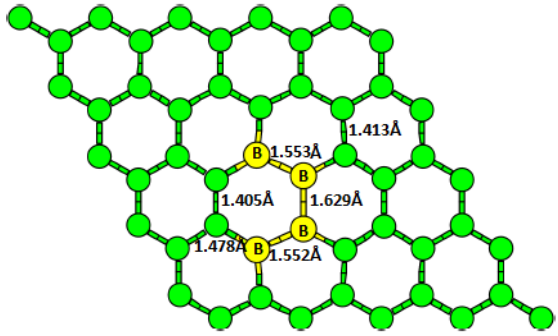
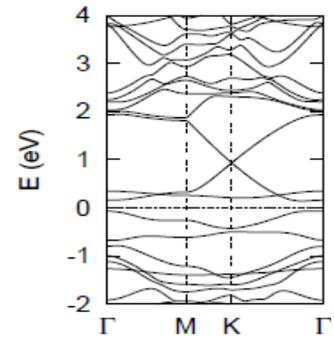

Fig. 6 (a)

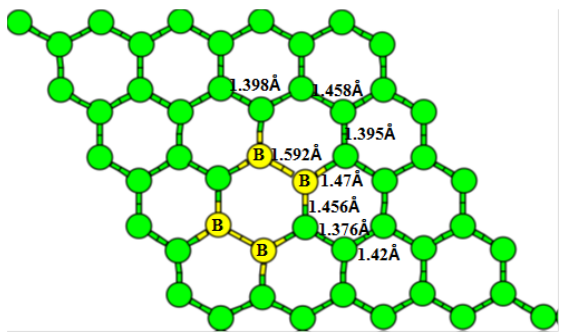
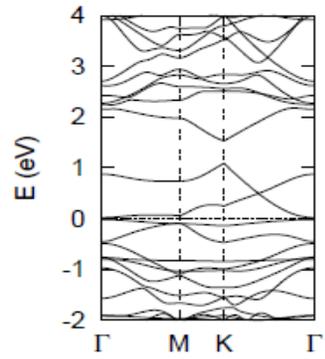

Fig. 6 (b)

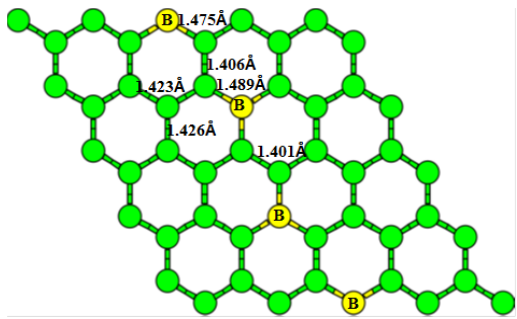
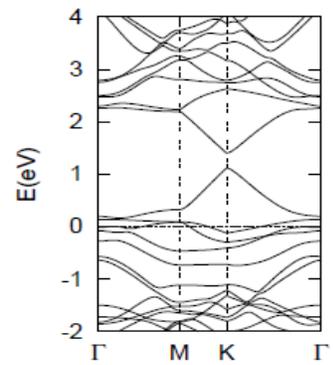

Fig. 6 (c)

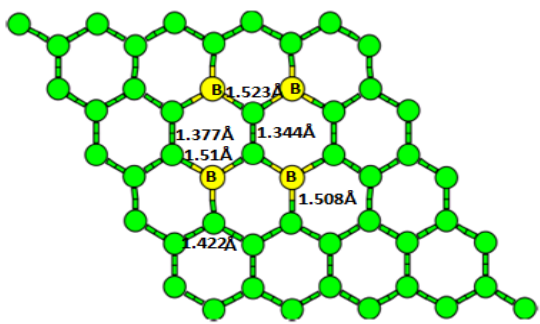
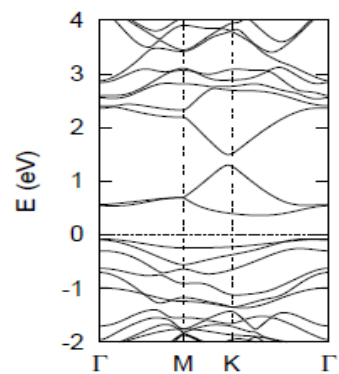

Fig. 6 (d)

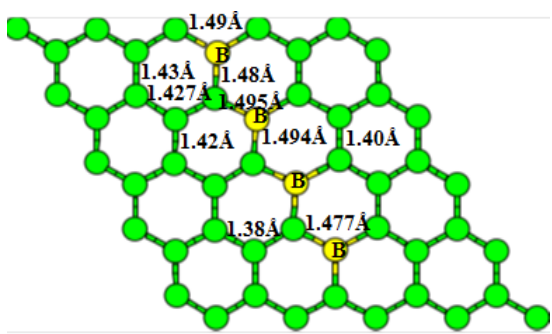
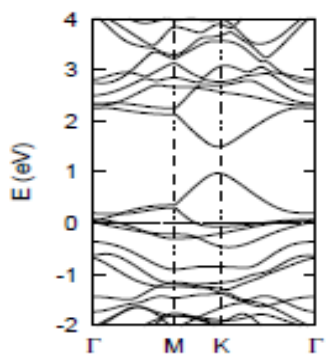

Fig. 6 (e)

**Fig.6  Some optimized geometries and band structures of four B atom doped graphene sheet**

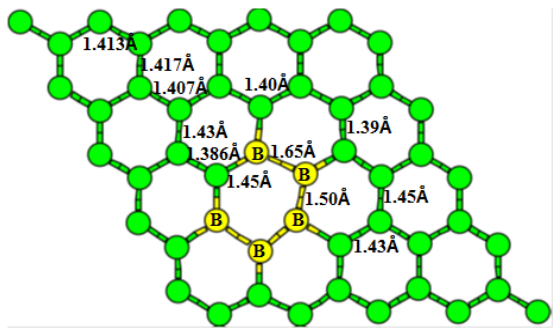
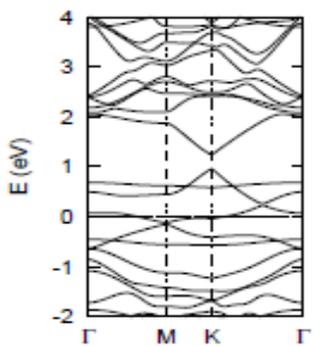

Fig. 7 (a)

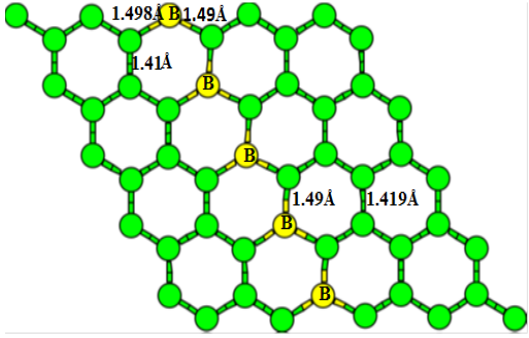 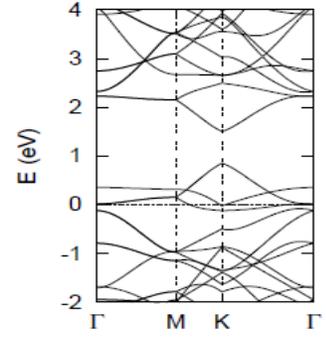

Fig. 7 (b)

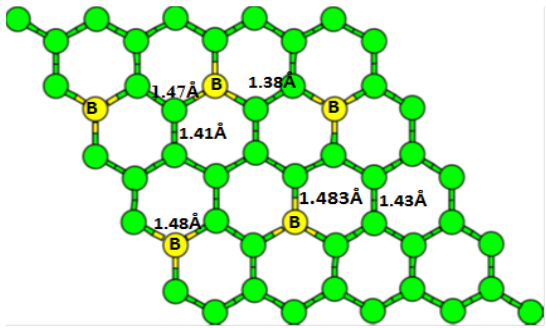 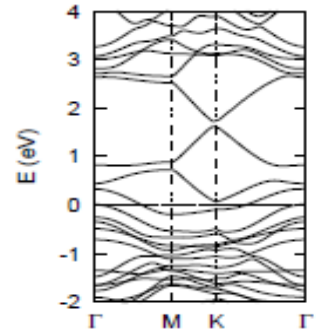

Fig. 7 (c)

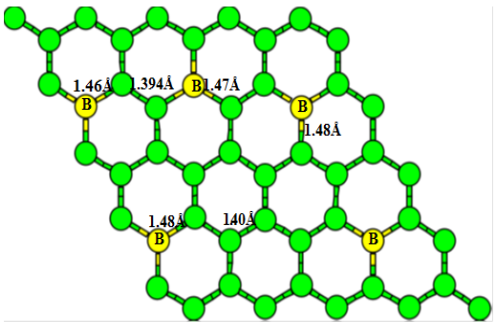 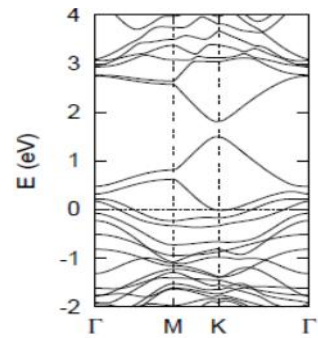

Fig. 7(d)

**Fig.7  Some optimized geometries and band structures of five B atom doped graphene sheet**

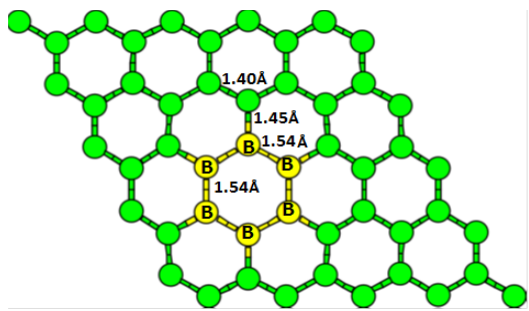 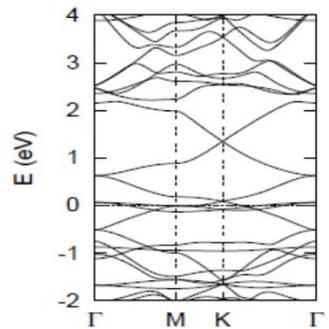

Fig. 8 (a)

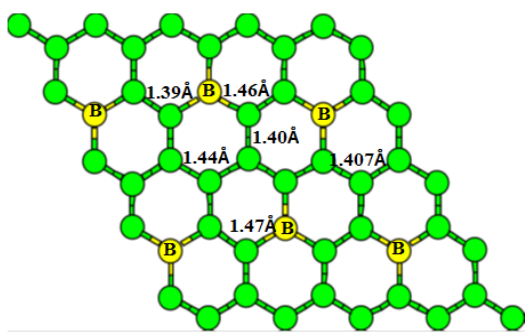 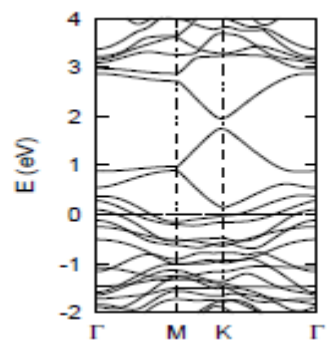

Fig. 8 (b)

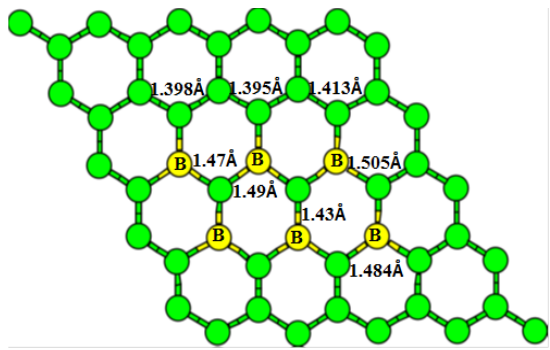 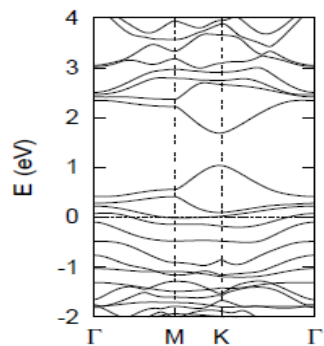

Fig. 8 (c)

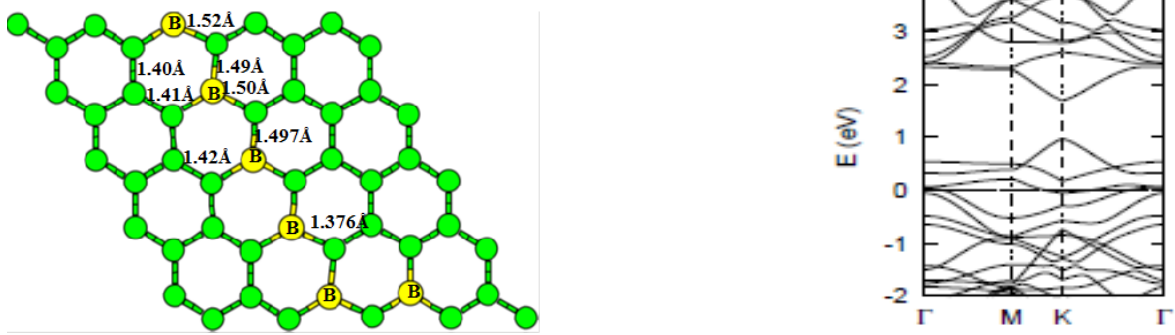

Fig. 8 (d)

**Fig.8 Optimized geometries and band structures of six boron atom doped graphene sheet**

The investigations of structural properties show the shrinkage of C-C bond length in case of B due to longer C-B bond because of large covalent radius of boron than carbon. This also results in disorder of the lattice when the dopants are placed at adjacent positions. The data regarding changes in bond lengths can be read from the structural part of the figures on the left are presented. It may be mentioned that the results regarding bond length alteration are in general consistent with earlier calculations of doping.

The electronic structure analysis shows an increase of band gap with concentration of doping. We give in Fig.9 a separate plot of this variation of band gap with % concentration of doped B atoms.

Now since the band gap results due to symmetry breaking of graphene sub-lattices, when the hetero-atoms are substituted at same sub-lattice points (A or B) for each concentration, the effect maximizes resulting into largest band gap. Also when the dopants are placed at adjacent positions the band gap is less for even number. of dopants than odd number of dopants ( band gap is completely closed in six atom doped in on hexagon). This is due to symmetry in the triangular sub-lattices formed by the hetero-atoms.

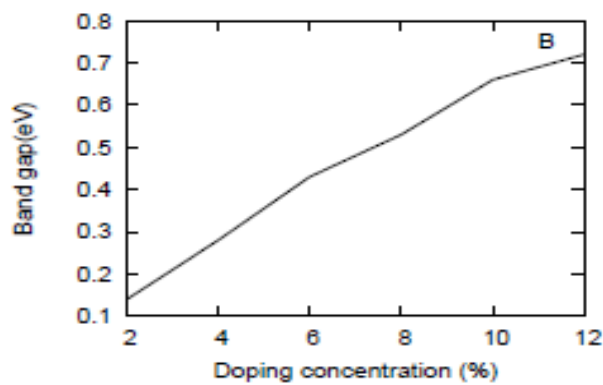

**Fig 9.  Band  gap in increasing order of  doping concentrations for boron doped graphene sheet. The isomers with highest band gap is chosen.**

## 3.2 Nitrogen doping

When the nitrogen atom is doped into the graphene sheet, similar effects as in case of Boron are observed. It also interacts through $sp^2$ hybridization. The bond length of three N-C bonds formed is 1.40 Å and there is almost no distortion in the planar structure of graphene. But due to electron rich character of the resulting structure, the Fermi level shifts by 0.7eV above the Dirac point. Also like boron the due to breaking of the symmetry of the graphene sub-lattices, the system shows a band gap of 0.14 eV. Bader charge analysis shows a charge transfer of 1.16 e from carbon to nitrogen [11]. Some data of single N atom doped graphene for comparison with other calculations is provided in Table 2. We present the optimized structure as well as the electronic band structure of some isomers obtained by doping increasing number of Nitrogen doped atoms in Figs.11-15.

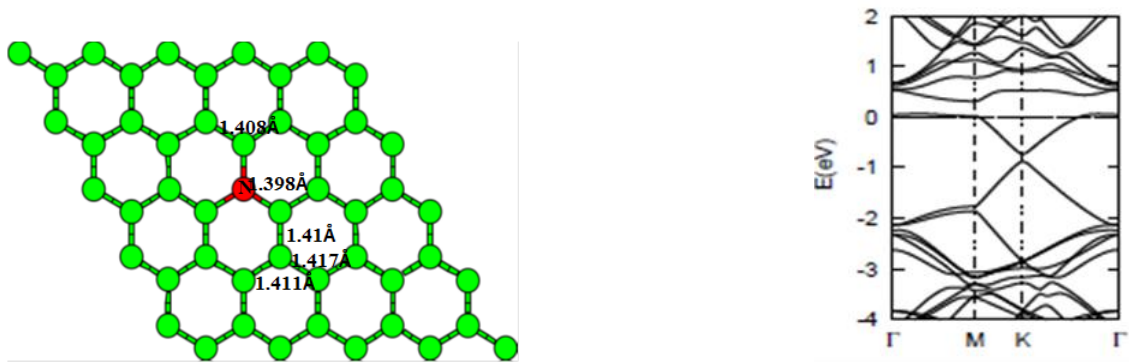

**Fig.10   Optimized geometry and band structures of nitrogen atom doped graphene sheet**

Table 2. Some parameters affecting single atom doping by Nitrogen atom

| Parameter | $d_{N-C}$ (Å) | Band gap (eV) | Charge Transfer(e) |
|-----------|---------------|---------------|--------------------|
| Our Work  | 1.40          | 0.14          | 1.16               |
| Ref.11    | 1.42          | 0.14          | 1.12               |

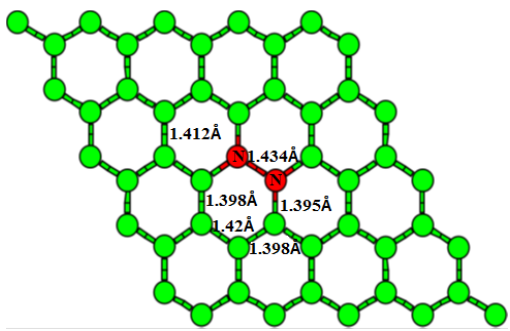 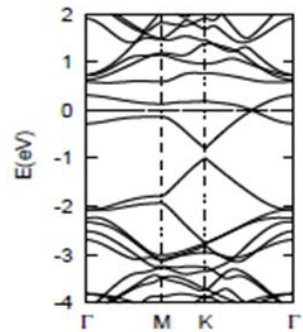

Fig. 11 (a)

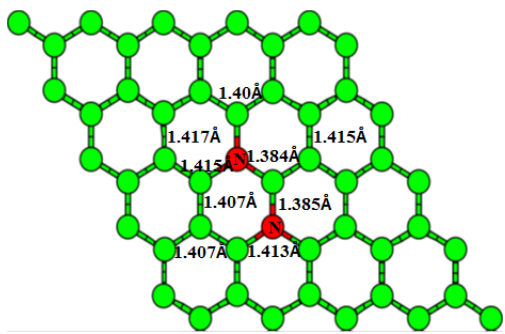 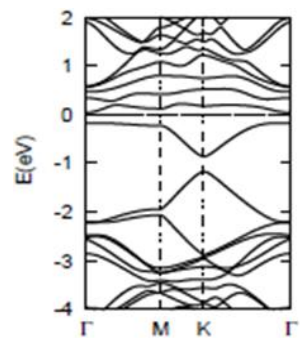

Fig. 11 (b)

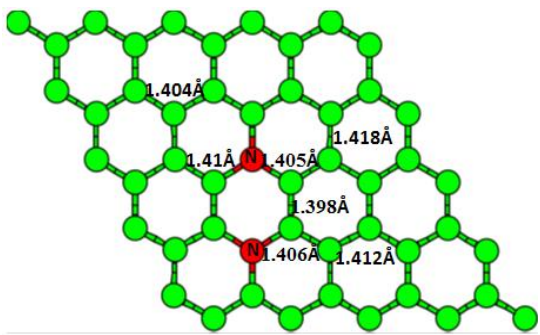 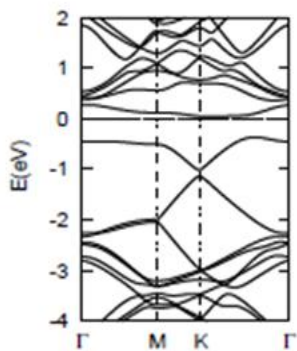

Fig. 11 (c)

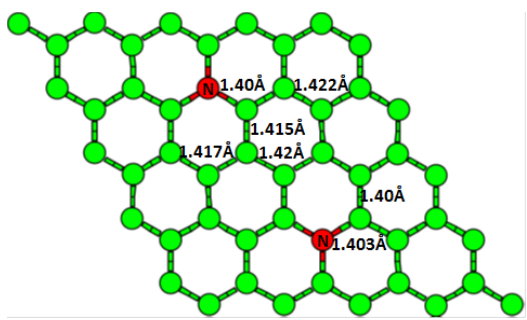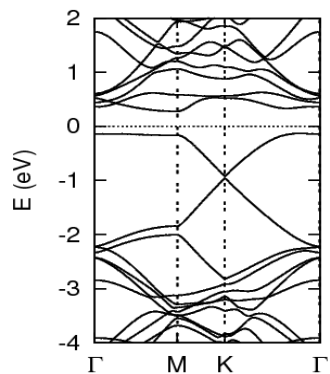

Fig. 11 (d)

**Fig.11 Optimized geometries and band structures of two nitrogen atom doped graphene sheet**

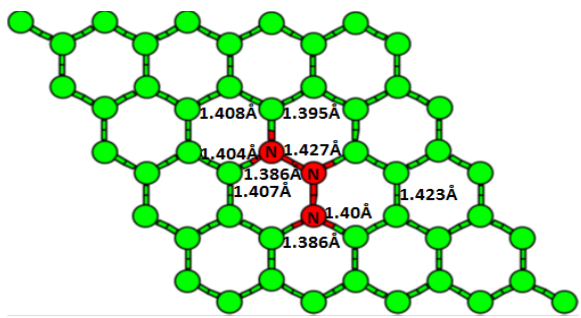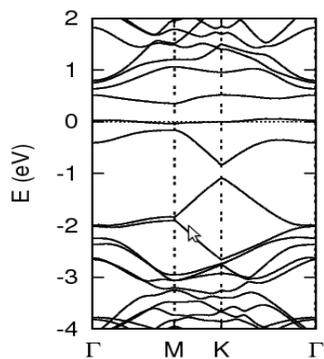

Fig. 12 (a)

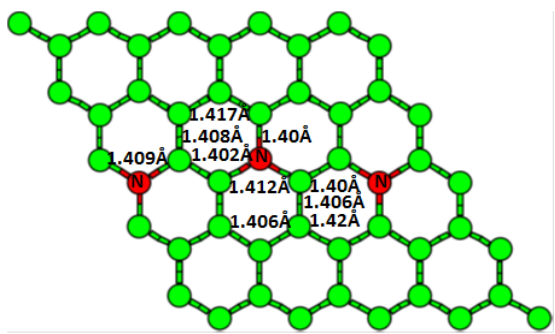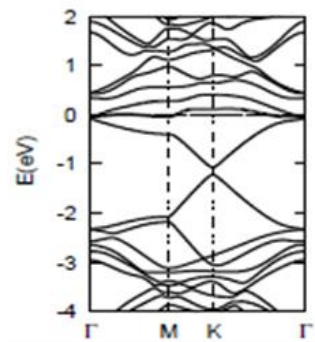

Fig. 12 (b)

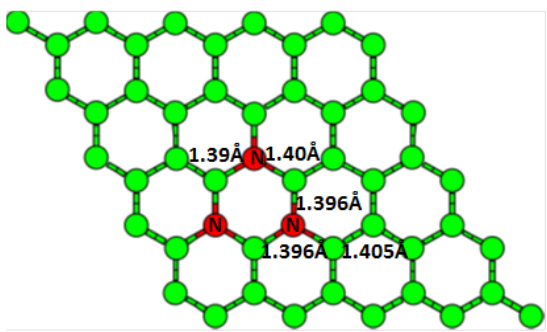 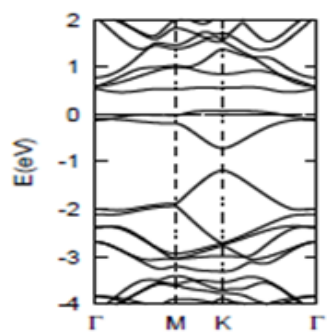

Fig. 12 (c)

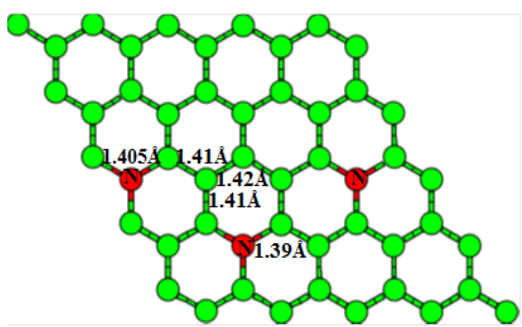 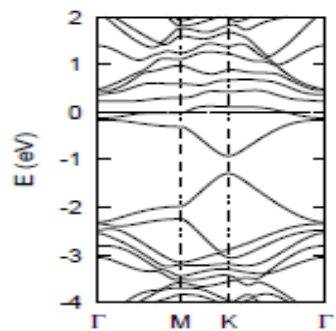

Fig. 12 (d)

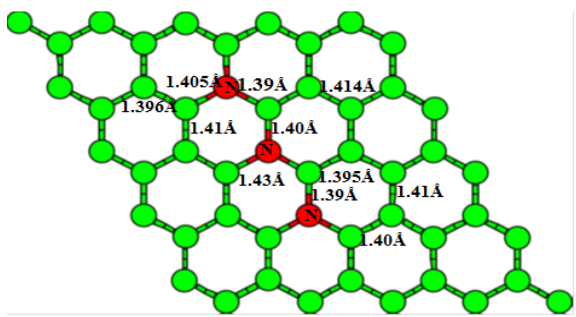 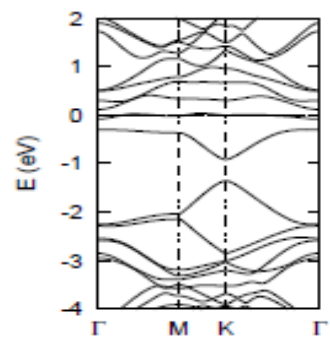

Fig. 12 (e)

**Fig.12 Optimized geometries and band structures of three nitrogen atom doped graphene sheet**

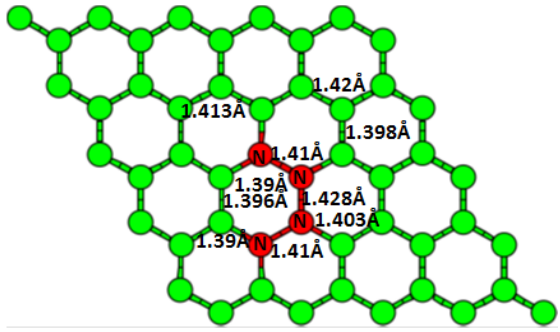
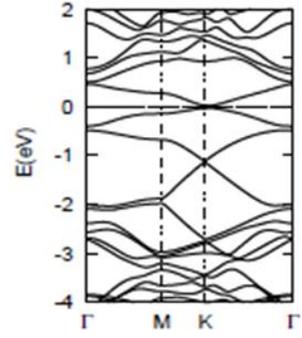

Fig. 13(a)

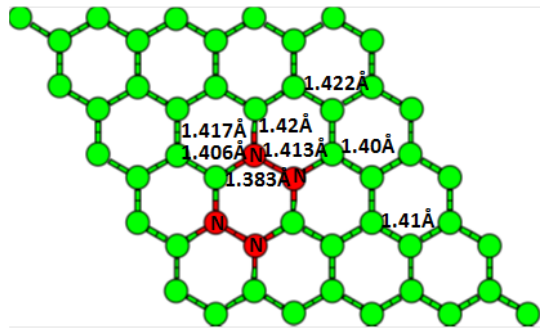
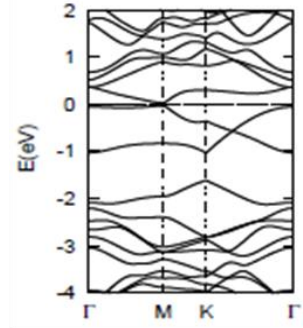

Fig. 13(b)

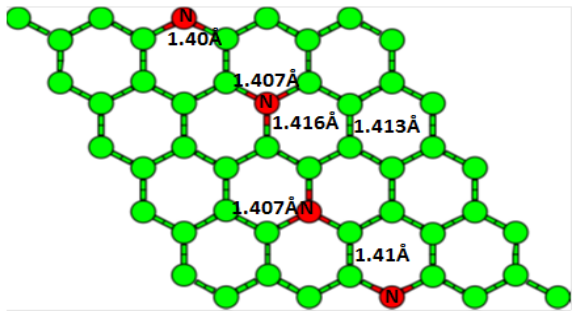
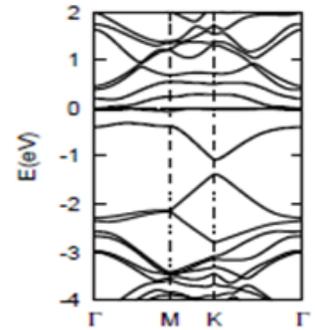

Fig. 13(c)

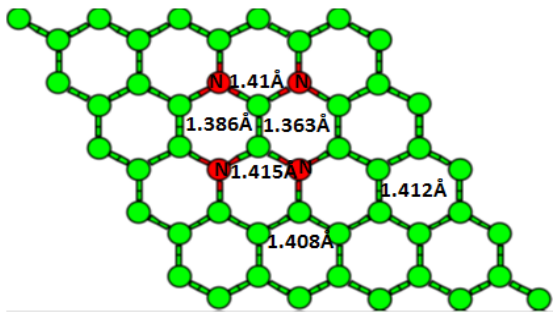 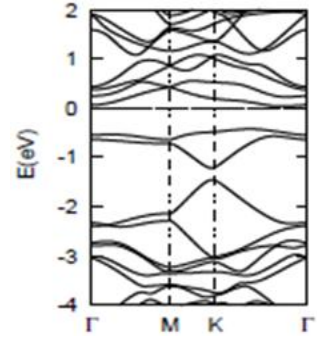

Fig. 13 (d)

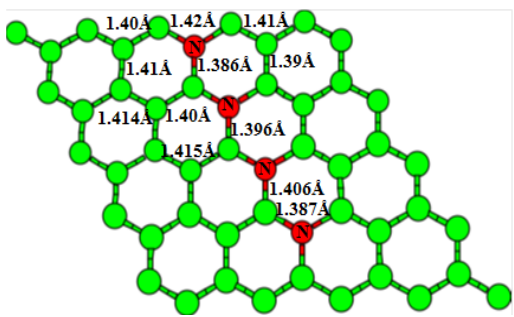 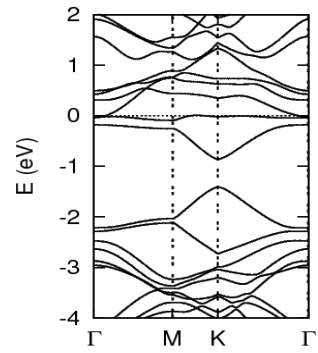

Fig. 13 (e)

**Fig.13   Some optimized geometries and band structures of four N atom doped graphene sheet**

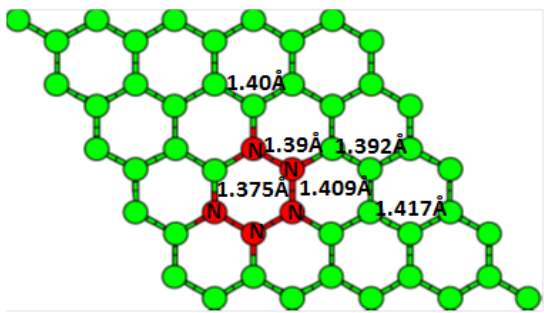 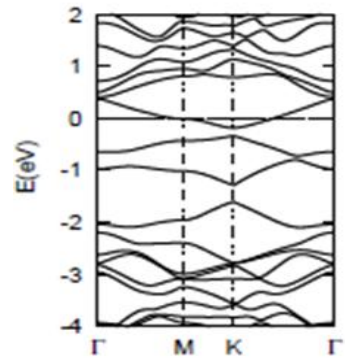

Fig. 14 (a

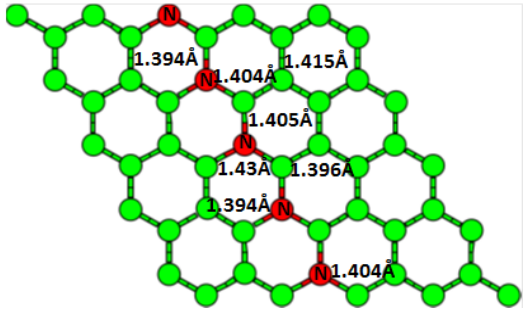
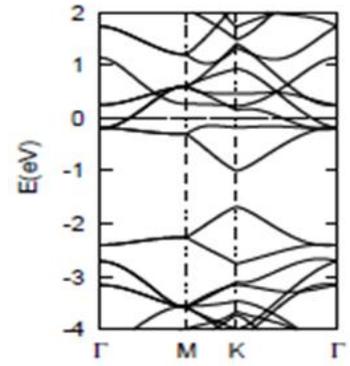

Fig. 14 (b)

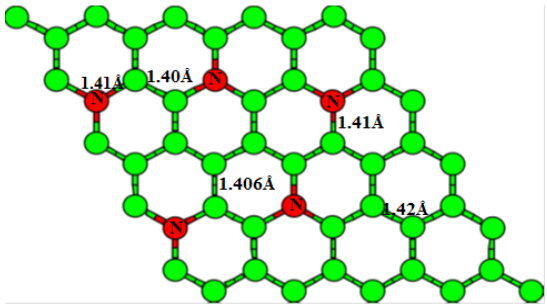
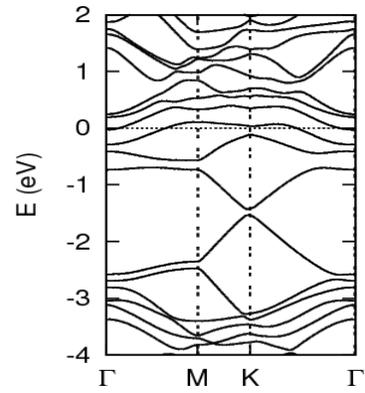

Fig. 14 (c)

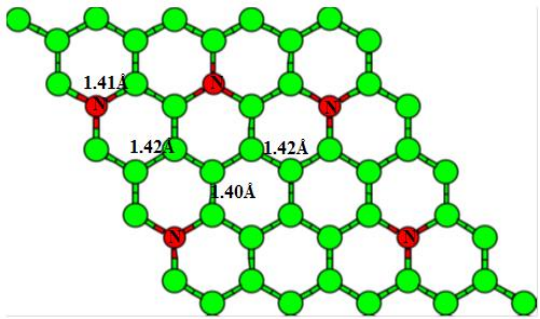
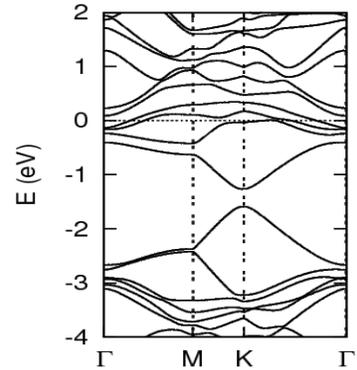

Fig. 14 (d)

**Fig.14 Optimized geometries and band structures of five nitrogen atom doped graphene sheet**

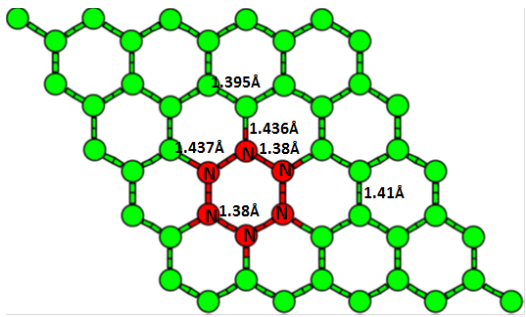
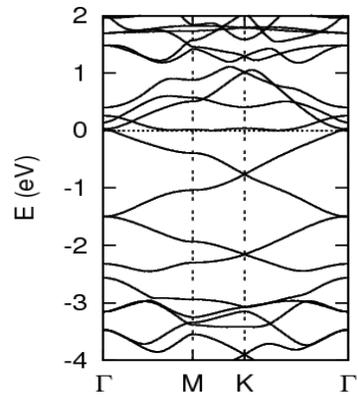

Fig. 15 (a)

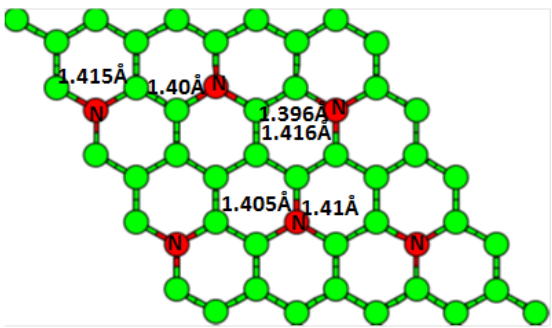
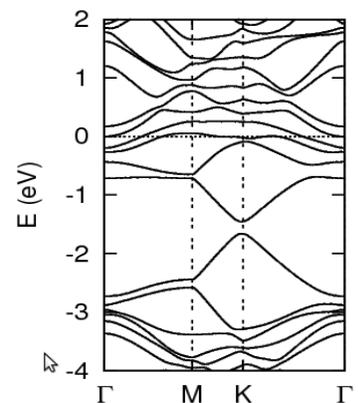

Fig. 15 (b)

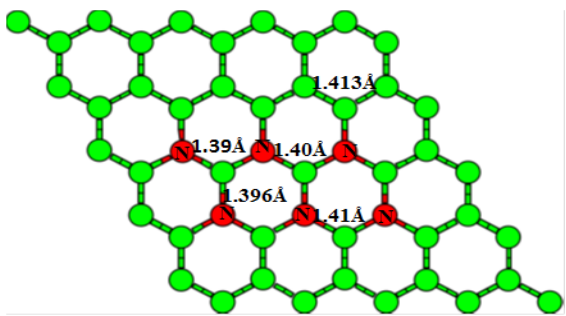
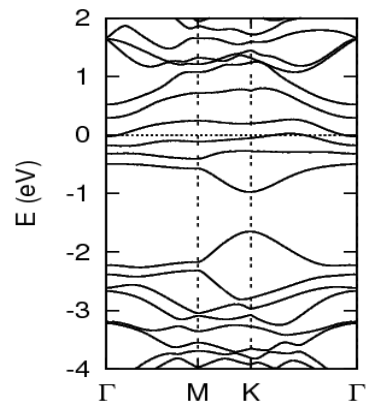

Fig. 15(c)

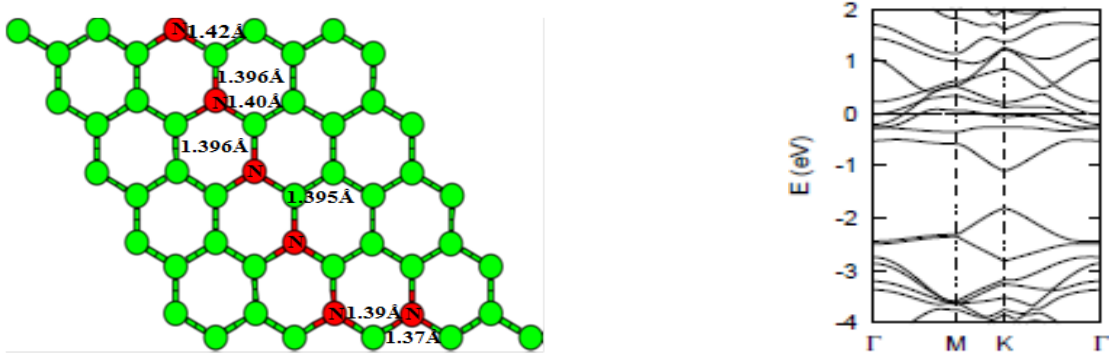

Fig. 15(d)

**Fig.15 Optimized geometries and band structures of six nitrogen atom doped graphene sheet**

   The lattice and band structure resulting from various number of dopants of N atom are shown in Figs.9-14. The structural properties of N-doped graphene geometries do not show any remarkable change because of nearly same size of covalent radius of carbon and nitrogen. The trend in electronic properties is similar to that of boron doped graphene. We summarize in Table3, the important results by doping both B and N atoms. A plot of variation of band gap with doping concentration appears in Fig. 15.

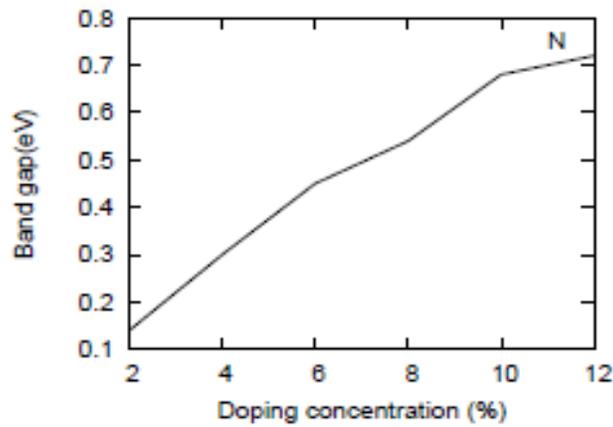

**Fig 16. Band gap in increasing order of doping concentrations ( for the configuration having maximum band gap) for N**

Table 3. Cohesive energy and Band gap for increasing doping concentration and various possible configurations a.b.c… as described in corresponding Figures.

| Concentration(%) | Configuration (B doped) | $E_{coh}^{*}$ (eV/atom) | Band gap(eV) | Configuration (N doped) | $E_{coh}^{*}$ (eV/atom) | Band gap(eV) |
|---|---|---|---|---|---|---|
| 2 | | -9.13 | 0.14 | | -9.16 | 0.14 |
| 4 | a | -9.03 | 0.15 | a | -9.10 | 0.19 |
| | b | -9.05 | 0.28 | b | -9.12 | 0.30 |
| | c | -9.06 | 0.14 | c | -9.13 | 0.15 |
| | d | -9.05 | 0.04 | d | -9.14 | 0.04 |
| 6 | a | -8.94 | 0.19 | a | -9.04 | 0.19 |
| | b | -8.98 | 0.17 | b | -9.08 | 0.16 |
| | c | -8.96 | 0.43 | c | -9.07 | 0.45 |
| | d | -8.97 | 0.34 | d | -9.09 | 0.36 |
| | e | -8.96 | 0.41 | e | -9.03 | 0.43 |
| 8 | a | -8.85 | 0.02 | a | -8.99 | 0.04 |
| | b | -8.86 | 0.46 | b | -9.01 | 0.50 |
| | c | -8.87 | 0.28 | c | -9.02 | 0.30 |
| | d | -8.89 | 0.23 | d | -9.05 | 0.29 |
| | e | -8.88 | 0.53 | e | -9.03 | 0.54 |
| 10 | a | -8.76 | 0.28 | a | -8.93 | 0.32 |
| | b | -8.81 | 0.66 | b | -8.99 | 0.68 |
| | c | -8.82 | 0.14 | c | -9.02 | 0.13 |
| | d | 8.81 | 0.32 | d | -9.00 | 0.33 |
| 12 | a | -8.64 | 0.00 | a | -8.84 | 0.003 |
| | b | -8.72 | 0.23 | b | -8.97 | 0.22 |
| | c | -8.71 | 0.65 | c | -8.95 | 0.68 |
| | d | -8.73 | 0.72 | d | -8.94 | 0.72 |

$$^{*}E_{coh}=[\ E_{tot}-n_i E_i\ ]/n \quad (i= C, B, N)$$

From several structural curves presented here for multi-atom B and N doped graphene, one notices a pattern of changes in some adjoining bond lengths with typical signature of Boron different from that of Nitrogen. Further, the band gaps continue to grow with increasing doping concentrations. The electronic band structure data has been summarized in table 3. In Table 3, $E_{coh}$ is the cohesive energy per atom of B- or N-doped configuration graphene and $E_{tot}$ and $E_i$ represent total energies of a structure and of individual elements present within the same supercell, respectively. $n_i$ is the number of $i^{th}$ species present in the configuration which could be C,B or N, while n is the total number of atoms present in a supercell (n = 50). However, with increasing concentration of dopants the cohesive energy decreases indicating the decreasing structural stability than the pristine graphene sheet ($E_{coh}$ = -9.20 eVper atom). Also the N-doped configurations show larger negative values of cohesive energy then their boron counterparts. This is due to nearly similar size of atomic radius of C and N which results in lesser structural disorder in comparison to B as is clear from bond length values.

# 4. CONCLUSIONS

Density functional theory was used to study the geometry and electronic structure of boron and nitrogen doped graphene sheet. The effect of doping was investigated by varying concentrations of dopants from 2 % (one atom of the dopant in 50 host atoms) to 12 % (six dopant atoms in 50 atoms host atoms) and also by considering different doping sites for the same concentration of substitutional doping. By B and N doping p-type and n-type doping is induced respectively in the graphene sheet. While the planar structure of the graphene sheet remains unaffected on doping, the electronic properties change from semimetal to semiconductor with increasing number of dopants. It has been observed that isomers formed by choosing various doping sites differ significantly in the stability, bond length and band gap introduced. The observation is that the positioning of dopant atoms also plays a significant role in modulating the band gap. The band gap is maximum when dopants are placed at same sublattice positions (A or B) in adjacent sublattices of graphene due to combined effect of symmetry breaking of sub lattices and the band gap is closed when dopants are placed at adjacent positions (alternate sublattice positions). These interesting results provide the possibility of designing the band gap of graphene as per requirements allowing its application in wide range of electronic devices. Indeed, it would not be out of place to mention that the role of N and B doping has also found favor [25] for replacing Platinum based cathode for polymer based electrolytic fuel cell (PEFC).

## ACKNOWLEDGMENTS

We express our gratitude to VASP team for providing the code, the computing facilities at IUAC (New Delhi) and the departmental computing facilities at Department of Physics, PU, Chandigarh. PR wishes to acknowledge UGC for financial support. VKJ acknowledges support from CSIR as Emeritus Scientist. He also acknowledges support of Alexander von Humboldt Foundation, Germany for extending research visit when this work got completed.

## REFERENCES

1.  K. S. Novoselov, A. K. Geim, S. V. Morozov and D. Jiang, Science 306, 666-669 (2004).

2.  A. K.  Geim and K. S. Novoselov, Nature Mater. 6, 183-191(2007).

3.  P. Avouris, Chen Z and V. Perebeinos  Nat. Nanotechnol. 2, 605 (2007).

4.  A. H. Castro Neto, F. Guinea, N. M. R. Peres, K. S. Novoselov and A. K.  Geim, Rev. Mod. Phys. 81, 109–162 (2009).

5.  Daniel R. Cooper, Benjamin D'Anjou, Nageswara Ghattamaneni, Benjamin Harack, Michael Hilke, Alexandre Horth, Norberto Majlis, Mathieu Massicotte, Leron Vandsburger, Eric Whiteway, and Victor Yu, ISRN Condensed Matter Physics, 2012,   501686 (2012).

6.  J. Dai, J. Yuan, P. Giannozzi, Appl. Phys. Lett. 95, 232105(3p) (2009).

7. Pablo A. Denis, Chem. Phy. Lett. 492, 251-257 (2010).

8. X. Wang, X. Li, Y. Yoon, P.K. Weber, H. Wang, J. Guo and H. Dai, Science 324,768-771 (2009) .

9. P. Shemella and S.K. Nayak, Appl. Phys. Lett. 94, 032101 (3pp) (2009).

10. A. Lherbie, R. X. Blasé, Y. Niquet, F. Triozon, F. & Roche, S. Phys. Rev. Lett. 101, 036808 (2008).

11. M. Wu, C. Cao and J. Z. Jiang, Nanotechnology 21, 505202(6pp) (2010).

12. E. H. Hwang, S. Adam and S. Das Sarma, Phys. Rev. B 76,195421 (6pp) (2007).

13. O. Leenaerts, B. Partoens, and F. M. Peeters,Phys. Rev. B 77, 125416 (6p) (2008).

14. H. Pinto, R Jones, J. P. Goss and P. R. Briddon, J. Phys.: Condens. Matter 21, 402001(3p) (2009).

15. Nabil Al-Aqtash,Khaldoun M. Al-Tarawneh,Tarek Tawalbeh and Igor Vasiliev, J. App. Phys. 112,034304 (2012).

16. Yingcai Fan, Mingwen Zhao,a Zhenhai Wang, Xuejuan Zhang, and Hongyu Zhang, App. Phys. Lett. 98,083103 (2011)

17. Xiaofeng Fan, Zexiang Shen, A. Q. Liu and Jer-Lai Kuo, Nanoscale 4, 2157 (2012).

18. Arun K. Manna and Swapan K. Pati, J. Phys. Chem. C, 115, 10842–10850 (2011).

19. G. Kresse and J. Furthmüller, Phys. Rev. B 54, 11169 (1996).

20. G. Kresse and J. Hafner, Phys. Rev. B, 47, 558 (1993).

21. W. Kohn and L. J. Sham, Phys. Rev. 140 A 1133-1138 (1965).

22. Perdew J P, Burke K and Ernzerhof, Phys. Rev. Lett. 77, 3865-3868 (1996).

23. P. E. Blöchl, O. Jepsend and O. K. Andersen, Phys. Rev. B, 49 16223 (1994).

24. A. L. E.Garcia, S. E. Baltazar, A.H.Romero, J. F. Perez Robelsand A. Rubio, Journal of Computational and Theoretical Nanoscience 5, 1-9 (2008).

25. Zhufeng Hou, Xianlong Wang, Takashi Ikeda, Kiyoyuki Terakura, Masaharu Oshima, Masa-aki Kakimoto, and Seizo Miyata, Phys. Rev. B **85**, 165439 (2012).